\documentclass[preprint,aps,superscriptaddress]{revtex4}
\usepackage[latin9]{luainputenc}
\usepackage{color}
\usepackage{array}
\usepackage{longtable}
\usepackage{multirow}
\usepackage{amsmath}
\usepackage{amssymb}
\usepackage{graphicx}
\usepackage{esint}
\usepackage[unicode=true,pdfusetitle,
 bookmarks=true,bookmarksnumbered=false,bookmarksopen=false,
 breaklinks=false,pdfborder={0 0 1},backref=false,colorlinks=false]
 {hyperref}

\makeatletter

\providecommand{\tabularnewline}{\\}

\@ifundefined{textcolor}{}
{%
 \definecolor{BLACK}{gray}{0}
 \definecolor{WHITE}{gray}{1}
 \definecolor{RED}{rgb}{1,0,0}
 \definecolor{GREEN}{rgb}{0,1,0}
 \definecolor{BLUE}{rgb}{0,0,1}
 \definecolor{CYAN}{cmyk}{1,0,0,0}
 \definecolor{MAGENTA}{cmyk}{0,1,0,0}
 \definecolor{YELLOW}{cmyk}{0,0,1,0}
}

\usepackage{amsfonts}
\providecommand{\U}[1]{\protect\rule{.1in}{.1in}}
\makeatother

\begin{document}
\title{Quantum Gravity Constraints on Fine Structure Constant from GUP in
Braneworlds }
\author{A. S. Lemos}
\email{adiellemos@gmail.com}

\affiliation{Departamento de F\'{i}sica, Universidade Federal de Campina Grande, Caixa
Postal 10071, 58429-900 Campina Grande, Para\'{i}ba, Brazil}
\author{F. A. Brito}
\email{fabrito@df.ufcg.edu.br }

\affiliation{Departamento de F\'{i}sica, Universidade Federal de Campina Grande, Caixa
Postal 10071, 58429-900 Campina Grande, Para\'{i}ba, Brazil}
\affiliation{Departamento de F\'{i}sica, Universidade Federal da Para\'{i}ba, Caixa Postal
5008, 58051-970 Jo\~{a}o Pessoa, Para\'{i}ba, Brazil}
\begin{abstract}
The Generalized Uncertainty Principle (GUP) has been discussed in
the thick braneworld scenario. By considering Rydberg atoms in this
background, we show that the spacetime geometry affects Maxwell equations
inducing an effective dielectric constant on the space. In its turn,
the corrected Coulomb potential by the gravitational interaction yields
a deviation on the $3$-dimensional Bohr radius. Then, we compute
the corrections on the fine structure constant owing to the GUP in
higher-dimensional spacetime. We also found constraints for the deformation
parameter $\beta$ and $D$-dimensional Planck length $l_{D}$ by
comparing the predicted deviations with the recent empirical data
of the fine structure constant. We compute the intermediate length
scale, which in principle may be larger than the Planck length scale.
It is conjectured that below such scale Quantum Gravity effects should
take place.
\end{abstract}
\keywords{GUP, braneworld, Rydberg states, fine structure constant}
\pacs{04.50.-h, 04.60.-m, 04.80.Cc, 32.10.Fn}
\maketitle

\section{Introduction}

There are theoretical motivations for the assumption of the high dimensionality
of spacetime. Originally this idea was presented by the schemes of
the well-known unification theories. Currently, its interest has been
renewed, mainly due to braneworld models. In this case, our observable
$4$-dimensional universe would be a thin hypersurface embedding in
a higher-dimensional space \citep{add1,add2,rs1,rs2}. In this scenario,
the particles and fields of the Standard Model (SM) would be trapped
on the brane, while gravity could have free access to the supplementary
space. The leakage of gravity to the extra dimensions would justify
the weakness of the gravitational interaction compared to the other
fundamental forces of nature and thus would solve the well-known hierarchy
problem \citep{add1,rs1}. From the field theory point of view, the
braneworld deals as a topological defect e.g., a domain wall \citep{RUBAKOV1983},
where the localization of fermions, for instance, can be ensured through
an interaction of the Yukawa-type between the Dirac fields and the
defect \citep{RUBAKOV1983}.

The well-known thick braneworld, which is an extension of the thin
brane models, has received increasing attention in the last years
\citep{CSAKI2000}. According to these models, the wave function of
the SM particles would have a finite penetration depth in the transverse
directions of the brane. Thus, the SM fields are confined on the submanifold
whose thickness in the extra dimensions should be of the order of
$\mathrm{\left(TeV/\hbar c\right)^{-1}}$, which would allow the emission
of SM particles into the bulk through collisions at $\mathrm{TeV}$
energies \citep{add1,add2}. 

Although the gravitational interaction is negligible in the atomic
domain according to $3$-dimensional physics, it will be amplified
in the extra-dimensional scenario. In this way, the atomic energy
spectrum is modified by the gravitational interaction. Thus, precise
measurements of atomic transitions can be used to test and constrain
extra-dimensional models. Indeed, the search for empirical traces
of extra dimensions in the thick brane scenario has motivated several
studies in atomic systems \citep{lemos1,lemos2,lemos3}. In this context,
we can highlight the recent work \citep{lemos3}, in which one computes
the gravitational correction on the energy levels of Hydrogen-like
ions that lies in Rydberg states. Though the gravitational interaction
is amplified in the short-distances regime, one has found that the
gravitational influence of the electrovacuum is the main contribution
to gravitational potential energy \citep{lemos3}.

While atomic spectroscopy has proven to be an alternative
way to seek traces of hidden dimensions, here, on the other hand,
we intend to discuss the possibility of measuring the effects of spatial
extra-dimensionality through the variations of the fine structure
constant in a thick brane scenario. Indeed, modern theories such as
brane models or string/M-theory offer a natural and self-consistent
framework for the variation of fundamental couplings \citep{PALMA2003,AGUILAR2003}.
Assuming that the multidimensional constants are genuinely fundamental,
they determine the couplings in three dimensions along with the size
$R$ of the space of extra dimensions. On its turn, in the thick brane
model, we expect that the brane structure leads to a deviation of
the $3$-dimensional value of the fine structure constant.

According to the quantum gravity theories -- such as string theory,
for instance -- it is expected that the strength of the gravitational
interaction will be comparable to other fundamental interactions on
the fundamental length scale that is assumably to be on the Planck
scale \citep{WILL2014} or electroweak scale in extra-dimensional
physics \citep{add1}. Since the absence of knowledge of how the physical
laws work on the Planck scale, also known as Planck length or minimal
length, it seems reasonable to seek to understand the theoretical
implications of the combined gravitational and quantum effects on
the properties of the physical systems in this fundamental scale.

Several theoretical models, such as the unification
schemes of string theory, black hole physics, and even doubly special
relativity (DSR) conjecture that, due to the existence of gravity,
the Heisenberg Uncertainty Principle (HUP) must be modified, which
in turn, leads to the prediction of the existence of a minimal observable
length \citep{AMATI1989,SCARDIGLI1999,ADLER1999,CAPOZZIELLO2000,CASADIO2023,GARRAY1995,AMELINO2001,ALI2009,HOSSEN2013}.
The extension of the HUP that encompasses these modifications is called
the Generalized Uncertainty Principle (GUP) \citep{TAWFIK2014}. Furthermore,
it is known that there are some derivations, and even proposals of
Gedanken experiments, generally associated with the gravitational
phenomena, which lead to the prediction of the GUP \citep{MAGGIORE1993}.
Thus, the GUP formulation incorporates gravitational corrections into
a principle of quantum mechanics.

Many works have employed the GUP formalism to the study of properties
and possible corrections to several physical systems such as BTZ black
hole \citep{ANACLETO2018,ANACLETO2021,IORIO2020} and acoustic black
hole \citep{ANACLETO2014}, approach to entropic nature of the gravitational
interaction \citep{NICOLINI2010}, study of the inflationary era \citep{TAWFIK2013},
description of brane cosmology dynamic \citep{BARCA2022}, and also
corrections for fine structure constant \citep{NASSERI2005,NASSERI2006,MARQUES2012}.
From a phenomenological point of view, experiments of gravitational
and non-gravitational sources have been investigated, aiming to obtain
constraints on the GUP deformation parameter \citep{SCARDIGLI2015,FENG2017,DAS2008,MARIN2013,BAWAJ2015,BUSHEV2019,GHOSH2014,DAS2009,ALI2011,BUONINFANTE2019,KANAZAWA2019,SCARDIGLI2020,SCARDIGLI2017,SCARDIGLI2019}.
In this perspective, it becomes important to study the phenomenology
of quantum gravity because we could infer measurable effects of physics
beyond the SM \citep{DAS2009}.

Given the recent developments in the field of spectroscopy
of Rydberg states \citep{JENTSCHURA2008}, some works have investigated
the possibility of testing for theories that lie beyond the Standard
Model from the study of the atomic transition between Rydberg states
\citep{lemos3,LEMOS2021,JONES2020}. Thus, this work aims to discuss
the constraints obtained from GUP in a thick brane scenario by considering
hydrogen-like atoms in Rydberg states. As we will see, the GUP introduces
the dependence on the higher-dimensional Planck length $l_{D}$ and
the $\beta$-parameter, called the deformation parameter, into the
corrected fine structure constant. The $\beta$-parameter estimates
the deformation on the canonical commutation relation from quantum
mechanics, and is directly related to the intermediate length scale
$\ell\equiv l_{D}\beta/2$ below which it is expected new physics
could come to play a fundamental role \citep{DAS2008}. We must also
worth highlight that, motivated by the extra-dimensional scenarios,
it seems reasonable to consider the Planck length as a parameter to
be constrained by the experiment. Then, we intend to constrain the
minimal length by comparing our theoretical prediction with the experimental
data from fine structure constant measurement. This analysis will
allow us, simultaneously, to constrain the deformation parameter $\beta$
from GUP.

The paper is organized as follows. In Sec. \ref{secII}, by considering
hydrogen-type atoms in Rydberg states, we show the potential produced
from the electromagnetic field generated by the atomic nucleus in
a thick braneworld scenario. For this case, the metric mimics the
Reissner-Nordstr\"{o}m geometry. Then we compute the gravitational correction
to the Coulomb electrostatic potential. In Sec. \ref{secIII}, we
address the GUP in the thick brane model and thus, obtain the corrected
fine structure constant. From this result, one find constraints on
higher-dimensional Planck length and deformation parameter by using
recent measurements of the fine structure constant. In its turn, in
Sec. \ref{new}, we compute the constraints on the deformation parameter
and the intermediate length scale, which must be a Quantum Gravity
effects regulator. Finally, in Sec. \ref{secIV}, we discuss some
results and present the final remarks.

\section{Rydberg Atom in the Thick brane\label{secII} }

We start by considering the proposed Arkani-Dimopoulos-Dvali (ADD)
braneworld model \citep{add1,add2}. In this case, the spacetime has
$\delta$ large extra dimensions compactified on a torus $T^{\delta}$
with a radius $R$. The SM fields must be localized in the $4$-dimensional
hypersurface by means of some confinement mechanism, while only the
gravitational field have access to extra-dimensional space \citep{add1}.
The classical action describing the gravitational field produced by
an atom in higher-dimensional space can be written as
\begin{equation}
S_{G}=\frac{c^{3}}{16\pi G_{D}}\intop d^{4}xd^{\delta}z\sqrt{-\hat{g}}\hat{R},\label{1}
\end{equation}
where $G_{D}$ is the higher-dimensional gravitational constant, $D=\delta+3$
labels the number of spatial dimensions, $\hat{g}$ is the determinant
of the metric with a signature $\left(-,+,\ldots,+\right)$ and $\hat{R}$
is the scalar curvature of the bulk. We should also mention that the
$x$-coordinates label the parallel directions to the brane while $z$-coordinates label the transverse directions.

In the atomic domain, we can consider gravity in the weak field regime
and thus write the metric as being composed of a four-dimensional
part and a perturbation term due to the existence of the supplementary
space. In this case, the metric assumes the form $g_{AB}=\eta_{AB}+h_{AB}$,
where $\eta_{AB}$ is the metric of Minkowski spacetime, and $h_{AB}$
is the perturbed metric that describes tensor fluctuations on flat
spacetime. Here, the atomic nucleus is responsible for generating
the geometry which resembles the Reissner-Nordstrom spacetime. Although
the charges are confined to $3$-brane and can not escape into the
bulk -- unless excited through highly energetic processes as in the
LHC experiment -- the electric field produced by the atomic nucleus
can penetrate a finite region $\varepsilon$ into the transverse directions.

By calculating the solution to the linearized Einstein equations in
a coordinate system that satisfies the harmonic gauge condition and
considering the appropriate energy-momentum tensor, it was shown that
the metric takes the form \citep{lemos3}:
\begin{eqnarray}
ds^{2} & = & -\left(1+\frac{2}{c^{2}}\varphi_{s}+\frac{2(2+\delta)}{c^{2}(1+\delta)}\chi_{s}\right)\left(dx^{0}\right)^{2}\nonumber \\
 & + & \left(1-\frac{2}{c^{2}(1+\delta)}\varphi_{s}\right)\left[\bigg(1+\frac{2(2+\delta)}{c^{2}(1+\delta)}\lambda_{1,s}\bigg)dr^{2}\right.\nonumber \\
 & + & \left.\bigg(1+\frac{2(2+\delta)}{c^{2}(1+\delta)}\lambda_{2,s}\bigg)r^{2}(d\theta^{2}+\sin^{2}\theta d\phi^{2})\right]\nonumber \\
 & + & \left(1-\frac{2}{c^{2}(1+\delta)}\varphi_{s}\right)d\vec{z}^{2},\label{metric}
\end{eqnarray}
where $\varphi_{s}$ is the Newtonian potential generated by an atom
while $\chi_{s}$ is the gravitational potential associated with the energy of the electromagnetic field. The functions $\lambda_{1,s}\cong\chi_{s}$,
and $\lambda_{2,s}\cong-\chi_{s}$ are found from spatial components
of the electromagnetic stress-energy tensor. For further details,
see \citep{lemos3}.

Recently, the potential $\varphi_{s}$ has been obtained for hydrogen-like
atoms in S-states by considering a Gaussian profile to nuclear wave-function
in the extra-dimensional directions \citep{lemos1,lemos2}. For this
case, one shows that the interior contribution of the gravitational
potential energy to atomic energy levels overcomes the outside contribution.
On the other hand, here we are interested in analyzing atoms with
a high principal quantum number known as Rydberg atoms. Therefore,
for the atoms which lie in Rydberg states the internal gravitational
energy contribution is suppressed by the external \citep{lemos3}.
In this case, the leading term of the gravitational potential can
be written as:
\begin{equation}
\varphi_{s}=-\frac{\hat{G}_{D}M}{r^{1+\delta}},\label{grav}
\end{equation}
where $\hat{G}_{D}=4G_{D}\Gamma(\frac{3+\delta}{2})/[(2+\delta)\pi^{\left(1+\delta\right)/2}]$.
This result is found by assuming that the Dirac delta approach to
describe the proton wave-function profile is valid. From \eqref{grav}
it can be shown that for Rydberg atoms on braneworld scenario, the
gravitational potential energy is amplified if $R\gg a_{0}$, where
$a_{0}$ is the Bohr radius.

Contrary to the gravitational potential $\varphi_{s}$, the gravitational
potential produced by the electromagnetic field, $\chi_{s}$, can
not be obtained in the thin brane limit by considering the delta distribution
to the atomic nucleus wave-function. Here the $\chi_{s}$-potential
diverges on all space due to the spread of the electromagnetic field
on $3$-brane. Therefore, for Rydberg atoms, we must admit that the
electric field flux lines have a penetration depth $\varepsilon\lesssim\mathcal{O}\left(10^{-19}\mathrm{m}\right)$
into the transversal directions to braneworld. In light of this fact,
one may calculate the gravitational potential inside the thick brane
due to the distribution of the electric field on spacetime and get
\citep{lemos3},
\begin{equation}
\chi_{s}=-\frac{\xi\left(\delta\right)}{16\pi\epsilon_{0}c^{2}}\frac{\hat{G}_{D}Q^{2}}{\varepsilon^{\delta-2}r^{4}},
\end{equation}
where $\varepsilon$ is the depth penetration of electric field lines
on the transverse direction to the brane, and $\xi\left(\delta\right)=\delta\sqrt{\pi}\Gamma\left(\frac{\delta-2}{2}\right)/2(\delta-1)\Gamma\left(\frac{\delta-1}{2}\right)$.
Note that,
\begin{equation}
\varphi_{s}/\chi_{s}=\frac{16\pi\epsilon_{0}c^{2}}{\xi\left(\delta\right)}\frac{M}{Q^{2}}\varepsilon\left(\frac{\varepsilon}{r}\right)^{\delta-3}\ll1,\;\text{if}\;\delta\geq3\;\text{and}\;r\geq r_{N},
\end{equation}
where $r_{N}$ is the nuclear radius. Its shown that $\chi_{s}$ surpass
the gravitational potential $\varphi_{s}$ for external regions to
the atomic nucleus by considering realistic values for the brane thickness
$\varepsilon<10^{-19}\mathrm{m}$. In the next section, we intend
to study the corrections induced by this spacetime geometry to $3$-dimensional
electrostatic potential. Thereby we expect to find corrections in
the electric potential due to extra-dimensional scenario.

\subsection{Gravitational Modified Maxwell's Equations }

Theoretical proposals for modifications of the electrostatic potential
have been presented in several scenarios. In string theory, for instance,
corrections to electrostatic potential were obtained and explained
due to the existence of the dilaton and modulus fields on effective
$4$-dimensional theory \citep{CVETIC1994}. On the other hand, from
a field theory point of view, modifications to Coulomb electric field
have been studied and associated with an exotic dielectric function
arising from the tachyon matter, which would correct Maxwell\textquoteright s
equations. It is shown that this approach mimics what occurs with
the field of gluons accountable for quarks binding in the hadronic
matter \citep{BRITO2014}.

In its turn, in our study, we expect that geometric effects associated
with the extra-dimensionality of spacetime might induce modifications
to electrostatic potential. Although the SM fields are localized on
the brane, the electric flux lines have a penetration depth $\varepsilon$
into transversal direction. However, at low energy scales, is expected
that the SM fields do not couple to the bulk geometry. Therefore,
without loss of generality, we can assume that the brane is located
at $z=0$, and so we write the metric induced into isotropic coordinates
as:
\begin{equation}
ds^{2}=-w^{2}\left(dx^{0}\right)^{2}+v^{2}\left(d\vec{x}\cdot d\vec{x}\right),
\end{equation}
where 
\begin{align}
w^{2} & =1+\frac{2}{c^{2}}\varphi_{s}+\frac{2(2+\delta)}{c^{2}(1+\delta)}\chi_{s},\label{metric_w}\\
v^{2} & =1-\frac{2}{c^{2}(1+\delta)}\varphi_{s}-\frac{(2+\delta)}{c^{2}(1+\delta)}\chi_{s}.\label{metric_v}
\end{align}

In curved spacetime, i.e., in the presence of gravity, the well-known
Maxwell\textquoteright s equation with source takes the form:
\begin{equation}
\frac{1}{\sqrt{-g}}\frac{\partial}{\partial x^{\mu}}(\sqrt{-g}F^{\mu\nu})=\mu_{0}J^{\nu},\label{maxwell}
\end{equation}
where $g$ is the determinant of the metric tensor, $F_{\mu\nu}=\partial_{\mu}A_{\nu}-\partial_{\nu}A_{\mu}$
is the field-strength tensor, $A_{\mu}$ is the four-potential, $J_{\nu}$
is the four-current and $\mu_{0}$ is the vacuum magnetic permeability.

We should note that since this problem is spherically symmetric due
to the presence of the static gravitational field produced by the
atomic nucleus, the equation of motion for the electric field reduces
to: 
\begin{equation}
\nabla\cdot(\epsilon\vec{E})=\rho_{0},\label{9}
\end{equation}
where $\epsilon=(v/w)\epsilon_{0}$ is an effective dielectric constant,
$\epsilon_{0}$ is the vacuum permittivity, and $\rho_{0}$ is the
charge density. Therefore, we observe that the effective dielectric
function $\epsilon$ modifies the electric field due to geometrical
effects. We must emphasize that, for conformally flat metrics whose
condition between the components of the metric $v=w$ is satisfied,
the $3$-dimensional Maxwell equation is recovered.

At low energy scales, we can write the electric field as being composed
of a 3-dimensional part corresponding to the Coulomb electric field
$\vec{E_{0}}$ and a correction owing to the gravitational contribution
$\vec{E_{g}}$. In this case, we admit the weak field limit, where
the gravitational interaction contributes with a small correction
to the electrostatic potential,
\begin{equation}
\phi=\phi_{0}+\phi_{g},
\end{equation}
where $\phi_{0}=Q/4\pi\epsilon_{0}r$ is the $3$-dimensional electrostatic
potential and $\phi_{g}\sim\ensuremath{\mathcal{O}}\left(\hat{G}_{D}\right)$
is the gravitational correction to the electrostatic potential, so
that the electric field assumes the form $\vec{E}=-\nabla\left(\phi_{0}+\phi_{g}\right)$.
In this case, the Eq. (\ref{9}) reduces to
\begin{equation}
\nabla^{2}\phi_{g}=-\frac{1}{\epsilon_{0}}\nabla\cdot\left[\left(1-\epsilon/\epsilon_{0}\right)\vec{E_{0}}\right].\label{eq:12}
\end{equation}
From the Eqs. (\ref{metric_w}) and (\ref{metric_v}) we get the solution
to Eq. (\ref{eq:12}), that will take the form:
\begin{equation}
\phi_{g}=\frac{\kappa\left(\delta\right)}{16\pi^{2}\epsilon_{0}^{2}c^{4}}\frac{\hat{G}_{D}Q^{3}}{r_{N}^{4}r\varepsilon^{\delta-2}}\left(1+\mathcal{O}\left(r_{N}/r\right)^{4}\right),
\end{equation}
where $\kappa\left(\delta\right)=3\sqrt{\pi}\delta(\delta+2)\Gamma\left(\frac{\delta-2}{2}\right)/64\Gamma\left(\frac{\delta+3}{2}\right)$,
$r_{N}=r_{0}A^{1/3}$ is the nuclear radius, where $A$ is the nucleon
number and $r_{0}=1.2\mathrm{fm}$ \citep{WONG2008}. Note that to
exterior region $r\gg r_{N}$ the contributions of higher orders will
be negligible since $\mathcal{O}\left(r_{N}/r\right)^{4}\ll1$.

\section{Fine structure constant from GUP in Higher-dimensional space\label{secIII}}

In this section, we obtain the corrections to the fine structure constant
owing to the GUP in the thick braneworld scenario. According to the
thick brane model, although the SM interactions are confined to $3$-dimensional
hypersurface, ensured by the existence of a higher-dimensional fundamental
Planck scale $M_{D}\sim\mathrm{\left(TeV/c^{2}\right)}$ \citep{add1,add2},
the wave-function of the SM particles have access to a limited region
into transverse directions to brane. As we have seen, by considering
a Rydberg atom in this background, a correction to the electrical
force experienced by charged particles orbiting the atomic nucleus
must arise and can be written as
\begin{equation}
F_{e}=\frac{Q^{2}}{4\pi\epsilon_{0}r^{2}}+\frac{\kappa(\delta)}{16\pi^{2}\epsilon_{0}^{2}c^{4}}\frac{\hat{G}_{D}Q^{4}}{r_{N}^{4}r^{2}\varepsilon^{\delta-2}}.\label{Fe}
\end{equation}
Note that if we turn off the gravitational effects present in the
second term of the Eq.(\ref{Fe}), we should recover the expected
classical result. From this relation, it is possible to calculate
the Bohr radius in the thick braneworld scenario. In this case, the
maximum uncertainty in the position of the lepton orbiting the atomic
nucleus of a hydrogen-like atom is related to the well-known Bohr
radius. So, for this higher-dimensional spacetime, we can easily obtain
the radius of the orbit of the nth state for a Rydberg atom:
\begin{equation}
\left(\Delta x_{i}\right)_{\mathrm{max}}\equiv\widetilde{a}_{0}=a_{0}n^{2}\left(1-\frac{a_{0}\hat{G}_{D}mQ^{4}\kappa(\delta)}{16\pi^{2}\epsilon_{0}^{2}c^{4}r_{N}^{4}\hbar^{2}\varepsilon^{\delta-2}}\right),\label{Dx}
\end{equation}
where $a_{0}=4\pi\epsilon_{0}\hbar^{2}/mQ^{2}$ is the $3$-dimensional
Bohr radius, $m$ is the lepton mass orbiting the atomic nucleus,
and $n$ is the principal quantum number. As seen, the extra dimensions
induce a correction to the $3$-dimensional Bohr radius.

\subsection{Fine structure constant corrected by GUP}

It is known that the limit to simultaneous measurement accuracy between
the position and momentum, for instance, is predicted by the HUP,
although there is no restriction for measuring the particle position.
On the other hand, the existence of a minimal measurable length has
been proposed in several contexts, including extra-dimensional scenarios,
and this leads to GUP \citep{TAWFIK2014}. Therefore, it is expected
that on the Planck scale -- that within the higher-dimensional framework
is $\sim\mathrm{\mathrm{\left(TeV/\hbar c\right)^{-1}}}$ -- the
effects of quantum gravity would manifest leading to correction to
the Heisenberg uncertainty relation. Following these ideas, the most
general GUP, including linear and quadratic contributions in momentum,
can be expressed as \citep{ANACLETO2021},
\begin{equation}
\Delta x_{i}\Delta p_{i}\geq\frac{\hbar}{2}\left(1-\beta\frac{l_{D}}{\hbar}\Delta p_{i}+\beta^{2}\frac{l_{D}^{2}}{\hbar^{2}}\Delta p_{i}^{2}\right),\label{GUP}
\end{equation}
where $l_{D}=\left(\hbar G_{D}/c^{3}\right)^{1/\left(\delta+2\right)}$
is the Planck length of the higher-dimensional spacetime, which from
braneworld point of view acts as a fundamental scale, $\beta$ is
a dimensionless positive deformation parameter. The commutation relation
that leads to the GUP (\ref{GUP}) and which is consistent with string
theory, DSR theories, and black holes physics has been originally
proposed in Refs.\citep{ALI2009,ALI2011,TAWFIK2015,GANGOPADHYAY2015,VAGENAS2019}.
Let us note that it is possible to construct an intermediate length
scale related to the minimal uncertainty on the position. Thereby,
we may rewrite the Eq. (\ref{GUP}) as follows:
\begin{equation}
\Delta x_{i}\geq\frac{\hbar}{2}\left(\frac{1}{\Delta p_{i}}-\beta\frac{l_{D}}{\hbar}+\beta^{2}\frac{l_{D}^{2}}{\hbar^{2}}\Delta p_{i}\right).\label{DeltaX}
\end{equation}
The minimum value to $\Delta x$ may be easily found by
\begin{equation}
\frac{d\left(\Delta x_{i}\right)}{d\left(\Delta p_{i}\right)}=\frac{\hbar}{2}\left(-\frac{1}{\Delta p_{i}^{2}}+\beta^{2}\frac{l_{D}^{2}}{\hbar^{2}}\right)=0
\end{equation}
Thus, the minimization of the Eq. (\ref{DeltaX}) is guaranteed for
$\left(\Delta p_{i}\right)_{\mathrm{max}}=\hbar/\beta l_{D}$, such
that now we can define an intermediate length scale bellow which we
expect that signals of new physics appear \citep{DAS2008},
\begin{equation}
\left(\Delta x_{i}\right)_{\mathrm{min}}=\ell\equiv\beta l_{D}/2.\label{Intermediate_Length}
\end{equation}
As we will see later, the existence of this new intermediate length
scale ensures that $\beta\thickapprox1$ and, in addition, effects
of Quantum Gravity can be measurable. 

Here we must also state that, without loss of generality, one has
assumed that the additional terms to uncertainty present on the Eq.
(\ref{GUP}) are due to gravitational interaction since the coupling
constant is the higher-dimensional gravitational constant $G_{D}$.
Finally, note that by a direct inspection from HUP, $\Delta x\Delta p\geq\hbar/2$,
and the GUP Eq. (\ref{GUP}), it is possible to define an effective
Planck constant
\begin{equation}
\hbar_{\mathrm{eff}}=\hbar\left(1-\beta\frac{l_{D}}{\hbar}\Delta p_{i}+\beta^{2}\frac{l_{D}^{2}}{\hbar^{2}}\Delta p_{i}^{2}\right).\label{h_eff}
\end{equation}

Now, let us analyze the solution for the momentum uncertainty from
Eq. (\ref{GUP}), aiming to determine the effective Planck constant
corrected due to the GUP in the braneworld framework. So, from Eq.
(\ref{GUP}), we find the uncertainty for the momentum as follows:
\begin{equation}
\Delta p_{i}=\frac{\hbar\left(\beta l_{D}+2\Delta x_{i}-\sqrt{4\Delta x_{i}^{2}+4\beta\Delta x_{i}l_{D}-3\beta^{2}l_{D}^{2}}\right)}{2\beta^{2}l_{D}^{2}}.\label{Dp}
\end{equation}

Since the Bohr radius can be related to the maximum uncertainty in
the position of lepton within the atom, we can substitute the results
(\ref{Dx}) and (\ref{Dp}) into Eq. (\ref{h_eff}), to obtain
\begin{eqnarray}
\hbar_{\mathrm{eff}} & \simeq & \hbar\left[1-\beta\left(\frac{l_{D}}{2a_{0}n^{2}}+\frac{\hat{G}_{D}l_{D}mQ^{4}\kappa\left(\delta\right)}{32\pi^{2}\epsilon_{0}^{2}c^{4}n^{2}\hbar^{2}r_{N}^{4}\varepsilon^{\delta-2}}\right)\right.\nonumber \\
 & + & \left.\beta^{2}\left(\frac{l_{D}^{2}}{2a_{0}^{2}n^{4}}+\frac{\hat{G}_{D}l_{D}^{2}mQ^{4}\kappa\left(\delta\right)}{16\pi^{2}\epsilon_{0}^{2}c^{4}n^{4}\hbar^{2}a_{0}r_{N}^{4}\varepsilon^{\delta-2}}\right)\right].\label{h_eff2}
\end{eqnarray}
This result has been found by considering a power series expansion
for $\hat{G}_{D}\ll1$ and $l_{D}\ll1$.

In the following, since we aim to find deviations to the 3-dimensional
fine structure constant $\alpha$, we will analyze the induced gravitational
corrections on the fine structure constant owing to the GUP. Thus,
we define an effective fine structure constant $\alpha_{\mathrm{eff}}$
from the effective Planck constant (\ref{h_eff2}) so that
\begin{equation}
\alpha_{\mathrm{eff}}=\frac{Q^{2}}{4\pi\epsilon_{0}\hbar_{\mathrm{eff}}c}.\label{alpha}
\end{equation}

Finally, substituting the result (\ref{h_eff2}) into Eq. (\ref{alpha}),
we get
\begin{eqnarray}
\alpha_{\mathrm{eff}} & \simeq & \frac{Q^{2}}{4\pi\epsilon_{0}\hbar c}\left[1+\beta\left(\frac{l_{D}}{2a_{0}n^{2}}+\frac{\hat{G}_{D}l_{D}mQ^{4}\kappa\left(\delta\right)}{32\pi^{2}\epsilon_{0}^{2}c^{4}n^{2}\hbar^{2}r_{N}^{4}\varepsilon^{\delta-2}}\right)\right.\nonumber \\
 &  & \left.-\beta^{2}\left(\frac{l_{D}^{2}}{4a_{0}^{2}n^{4}}+\frac{\hat{G}_{D}l_{D}^{2}mQ^{4}\kappa\left(\delta\right)}{32\pi^{2}\epsilon_{0}^{2}c^{4}n^{4}\hbar^{2}a_{0}r_{N}^{4}\varepsilon^{\delta-2}}\right)\right].\label{alpha_eff}
\end{eqnarray}
Here, again, we have performed a power series expansion by assuming
that $\hat{G}_{D}\ll1$ and $l_{D}\ll1$. As seen, due to geometrical
effects, we find that the effective fine structure constant in higher-dimensional
spacetime presents deviations compared to its expected classical $3$-dimensional
value. Although $\alpha$ can be assumed as a ``coupling constant"
of the electromagnetic interaction, the gravitational field induced
corrections on their value extracted from Rydberg atoms in the thick
brane scenario. On its turn, in the absence of extra-dimensional effects,
i.e., for $\delta=0$ the $l_{D}$ reduces to $3$-dimensional Planck
length $l_{Pl}$, and so we obtain the fine structure constant corrected
by $3$-dimensional GUP, which for an hydrogen-like atom with $Z=1$
in the fundamental state can be written as follows:
\begin{equation}
\alpha_{eff}\simeq\alpha_{0}\left(1+1.53\times10^{-25}\beta-2.33\times10^{-50}\beta^{2}\right),
\end{equation}
where $\alpha_{0}=Q^{2}/4\pi\epsilon_{0}\hbar c$ is the $3$-dimensional
fine structure constant. This small correction found shows us that
the fine structure constant has a value very close to the $3$-dimensional
expected value.

\begin{figure}[th]
\includegraphics[scale=0.3]{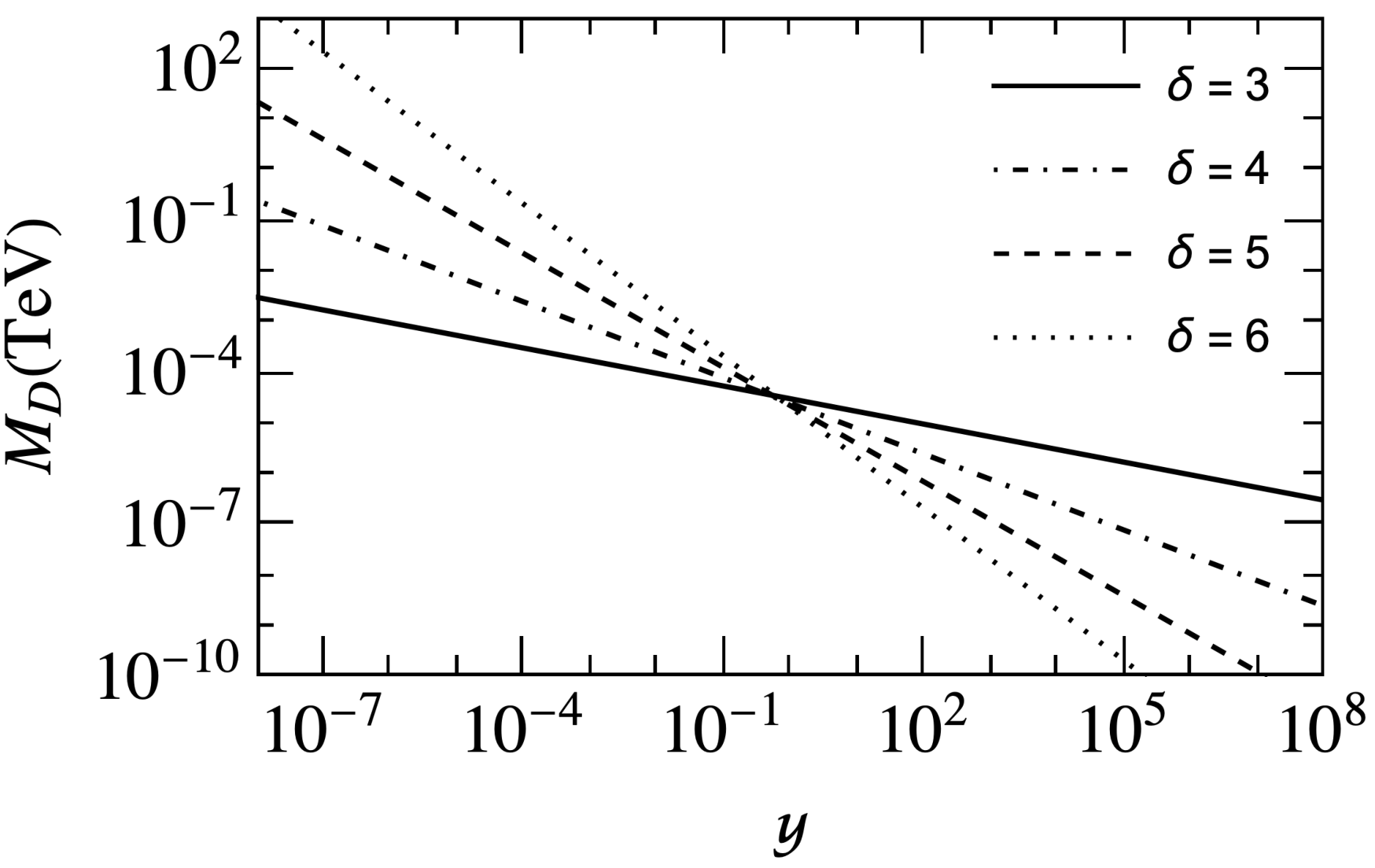}

\caption{\label{fig1}Requirements on higher-dimensional Planck mass so that
the correction from the electrovacuum exceeds the contribution from
$\ensuremath{\mathcal{O}}(\hat{G}_{D})^{0}$, by considering a hydrogen-like
Rydberg atom.}
\end{figure}

On the other hand, if we assume the condition predicted by the string
theory such that $\beta\ll1$, and since as $l_{D}/a_{0}\ll1$, one
can see from Eq. (\ref{alpha_eff}) that the first-order correction
in $\beta$ will be stronger than the second-order $\sim\beta^{2}$.
In this case, we can determine the regime for which the extra-dimensional
contribution owing of the brane structure for atoms in Rydberg states
will surpass the term arising from the generalization of the HUP,
i.e., the term of $\ensuremath{\mathcal{O}}\left(l_{D}\right)$. Thus,
we get the condition on the higher-dimensional Planck mass
\begin{equation}
M_{D}<\left[\frac{16\Gamma\left(\frac{\delta+3}{2}\right)\kappa(\delta)}{\left(\delta+2\right)\pi^{\left(\delta+1\right)/2}}\frac{m_{e}a_{0}}{\varepsilon^{\delta-2}r_{N}^{2}}\frac{\hbar^{\delta-1}}{c^{\delta+3}}\left(\frac{Q^{2}}{8\pi\epsilon_{0}r_{N}}\right)^{2}\right]^{1/\left(\delta+2\right)}.\label{MD}
\end{equation}
The condition (\ref{MD}) is shown in Fig. \ref{fig1}
and presents bounds on $M_{D}$ so that the contribution due to the
brane structure overcomes that one owing the term $l_{D}/2a_{0}n^{2}$.
For this purpose, let us admit that $l_{D}$ is related to $\varepsilon$
by a factor $y$, such that $\varepsilon=yl_{D}$.

From Figure \ref{fig1}, we should note that the curves represent
the upper limit for which the condition Eq. (\ref{MD}) is valid.
For values of $M_{D}$ in the area, below the curves, the gravitational
correction from electrovacuum overcomes the contribution due to the
first term $l_{D}/2a_{0}n^{2}$, i.e., the effects from thickness
braneworld should amplify the correction on the fine structure constant. In this case, for small values of the $y$-parameter,
the brane structure deviation will dominate fine structure correction,
while higher values imply the weakening of such correction. As we
will see in the following subsection, strict corrections coming from
$3$-dimensional GUP are smaller than uncertainty in the measure of
$\alpha$. In its turn, we expect that the extra-dimensional scenario
amplifies such corrections.

\subsection{Constraints for the GUP deformation parameter $\beta$ in braneworld
scenario}

In this subsection, we compute the bounds on the fine structure constant
owing to the corrections from the GUP in the thick brane scenario.
As proposed for theoretical models of quantum gravity, especially
in some string theory inspired models, the deformation parameter $\beta$
present on GUP must be of order unity \citep{BUONINFANTE2019}. On
the other hand, the empirical constraints owing to experiments of
gravitational and non-gravitational sources have been investigated,
aiming to obtain upper bounds on the deformation parameter. In this
case, the deformation parameter $\beta$ can assume high values ($\beta\gg1$)
without such results implying conflicts with the empirical data \citep{KANAZAWA2019}.
Here, in order to present simultaneously constraints on the parameter
$\beta$ and $l_{D}$, let us analyze recent measurements of the fine
structure constant.

A recent measurement of the fine structure constant by using matter-wave
interferometry technique with rubidium atoms has been obtained with
an unprecedented precision \citep{MOREL2020},
\begin{equation}
\alpha^{-1}=137.035999206\left(11\right),\label{alpha_value}
\end{equation}
whose experimental uncertainty $\delta\alpha_{\mathrm{\mathrm{exp}}}$
is expressed in parenthesis. As we have seen, the theoretical GUP
model at the presence of a supplementary space predicts deviations
on the fine structure constant. So, from the measure (\ref{alpha_value}),
we intend to obtain bounds to proposed corrections. Firstly, from
Eq. (\ref{alpha_eff}), one can rewrite the effective fine structure
constant as
\begin{equation}
\alpha_{\mathrm{eff}}=\alpha_{0}+\Delta\alpha,
\end{equation}
where $\Delta\alpha$ is a correction due to GUP into thick brane
framework. Aiming to infer empirical bounds, we must require that
$\left|\Delta\alpha\right|<\delta\alpha_{\mathrm{\mathrm{exp}}}$,
which demands that the correction obtained does not exceed the uncertainty
in the measurement of the fine structure constant. In this way, we
ensure that, so far, no indications that confirm the existence of
a fundamental length of the large scale proposed by GUP or even extra
dimensions have been detected. 

\begin{figure}[th]
\begin{minipage}[c][1\totalheight][t]{0.45\textwidth}%
\begin{center}
\includegraphics[scale=0.4]{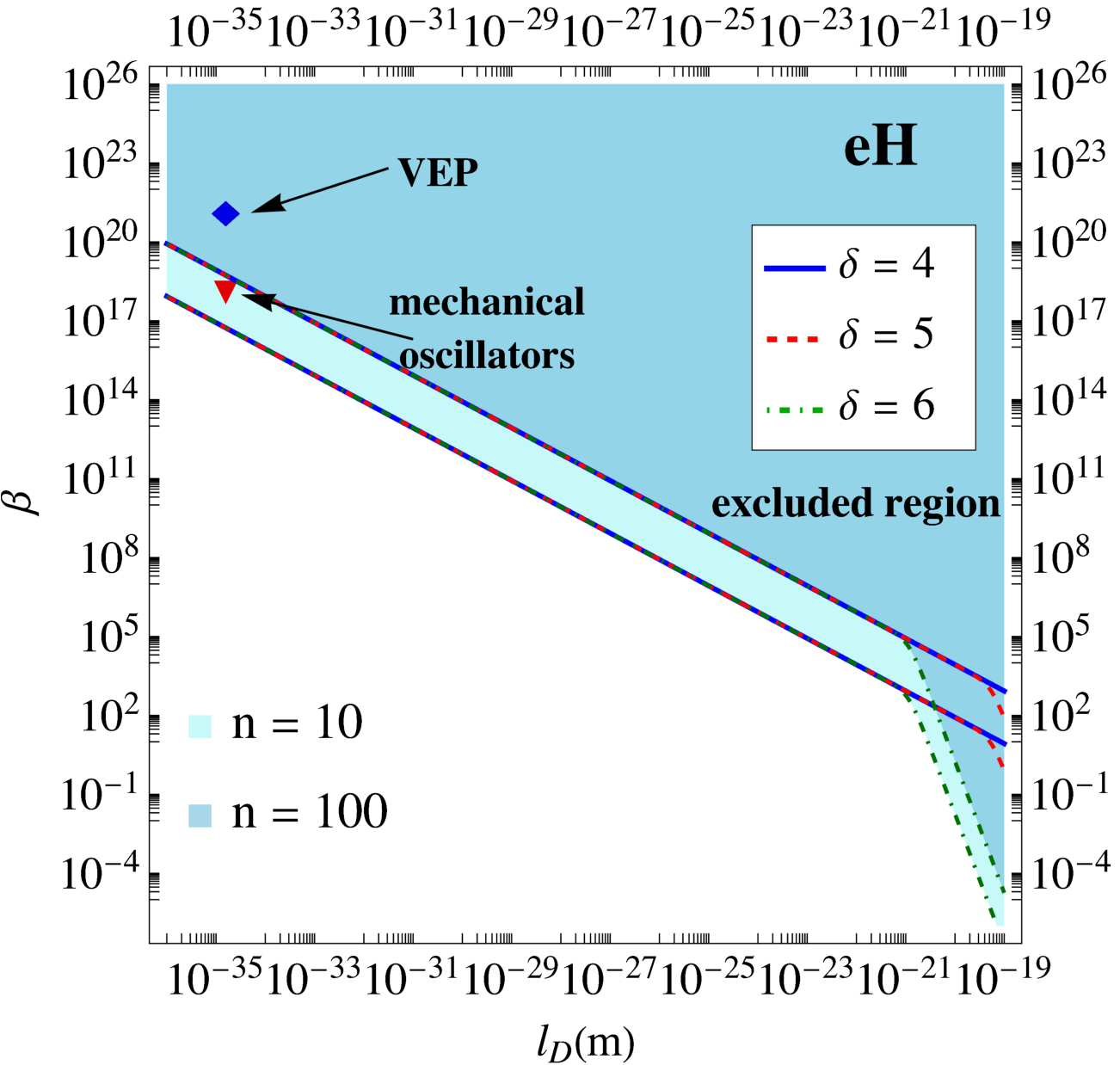}
\par\end{center}
\begin{center}
(a)
\par\end{center}%
\end{minipage}\hfill{}%
\begin{minipage}[c][1\totalheight][t]{0.45\textwidth}%
\begin{center}
\includegraphics[scale=0.4]{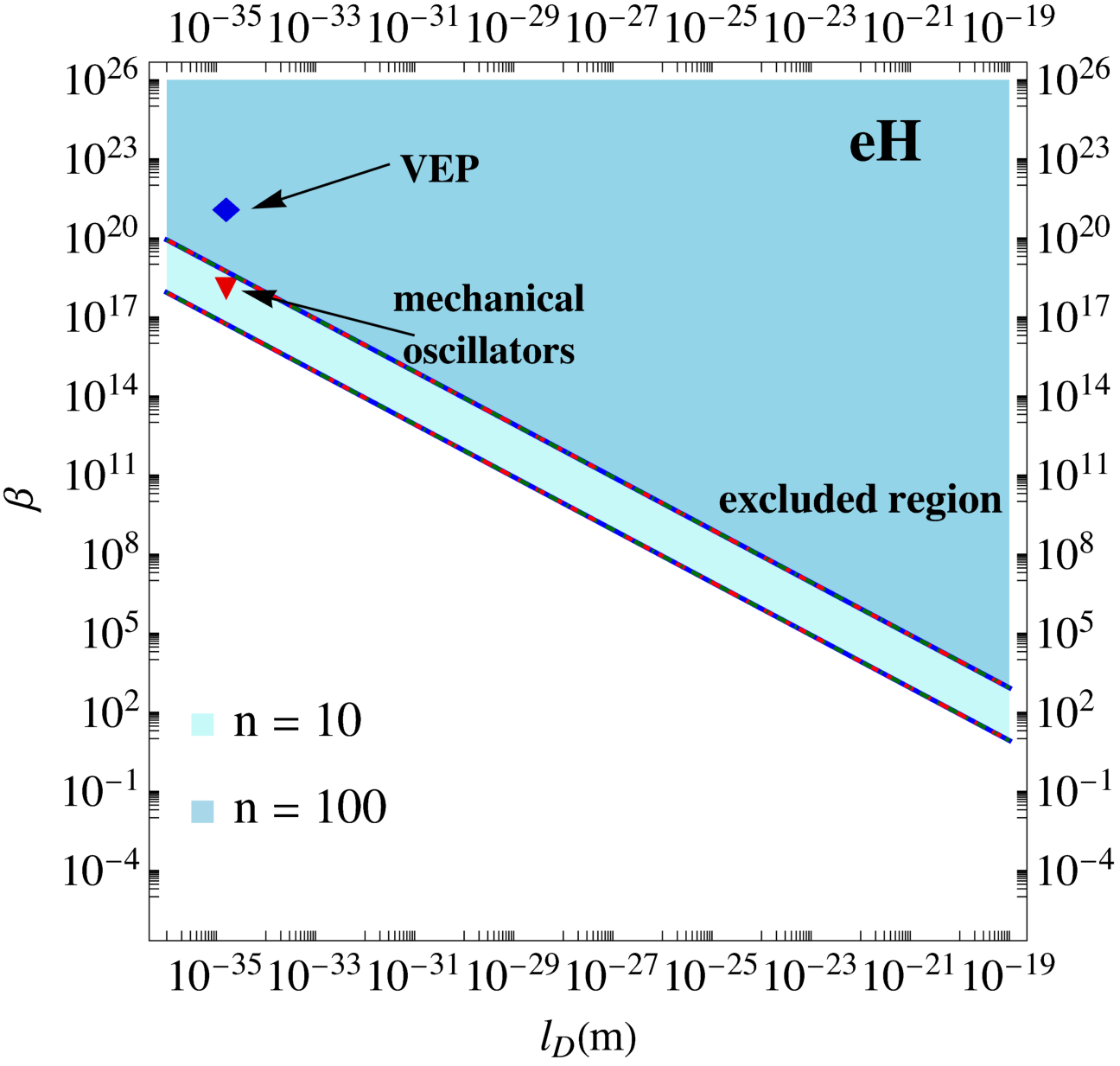}
\par\end{center}
\begin{center}
(b)
\par\end{center}%
\end{minipage}\caption{\label{fig2}Constraints are shown to corrections of the fine structure
constant induced from the $\beta$ deforming parameter owing to GUP
into higher-dimensional spacetime by considering electronic hydrogen-like
atoms in Rydberg states. The symbols ``$\blacklozenge$" and ``$\blacktriangledown$"
labels respectively bounds from gravitational and non-gravitational
empirical data \citep{KANAZAWA2019}. In Fig. (a),
we have assumed $y=10^{-6}$, while in Fig. (b) $y=10^{6}$ was considered.}
\end{figure}

\begin{figure}[th]
\begin{minipage}[c][1\totalheight][t]{0.45\textwidth}%
\begin{center}
\includegraphics[scale=0.4]{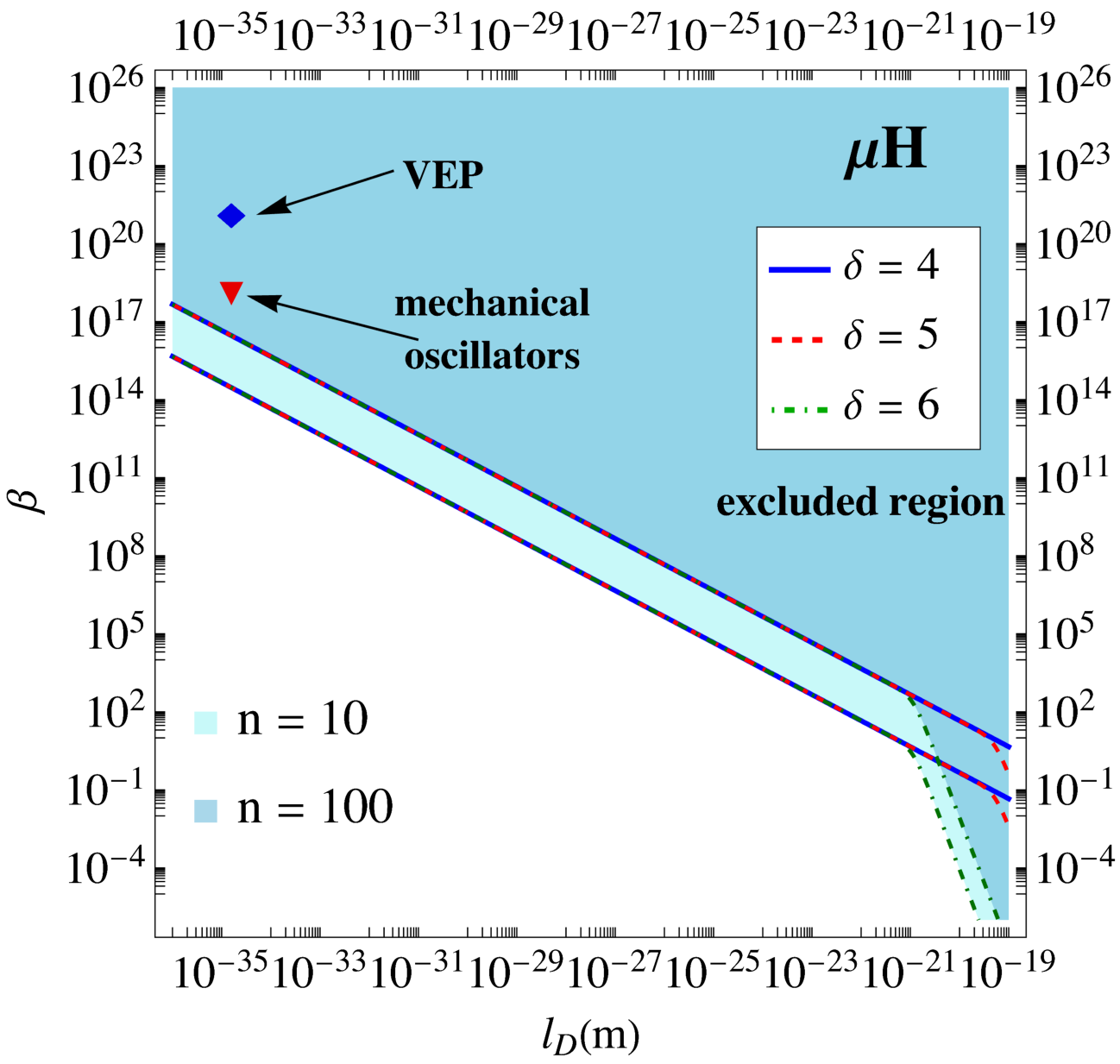}
\par\end{center}
\begin{center}
(a)
\par\end{center}%
\end{minipage}\hfill{}%
\begin{minipage}[c][1\totalheight][t]{0.45\textwidth}%
\begin{center}
\includegraphics[scale=0.4]{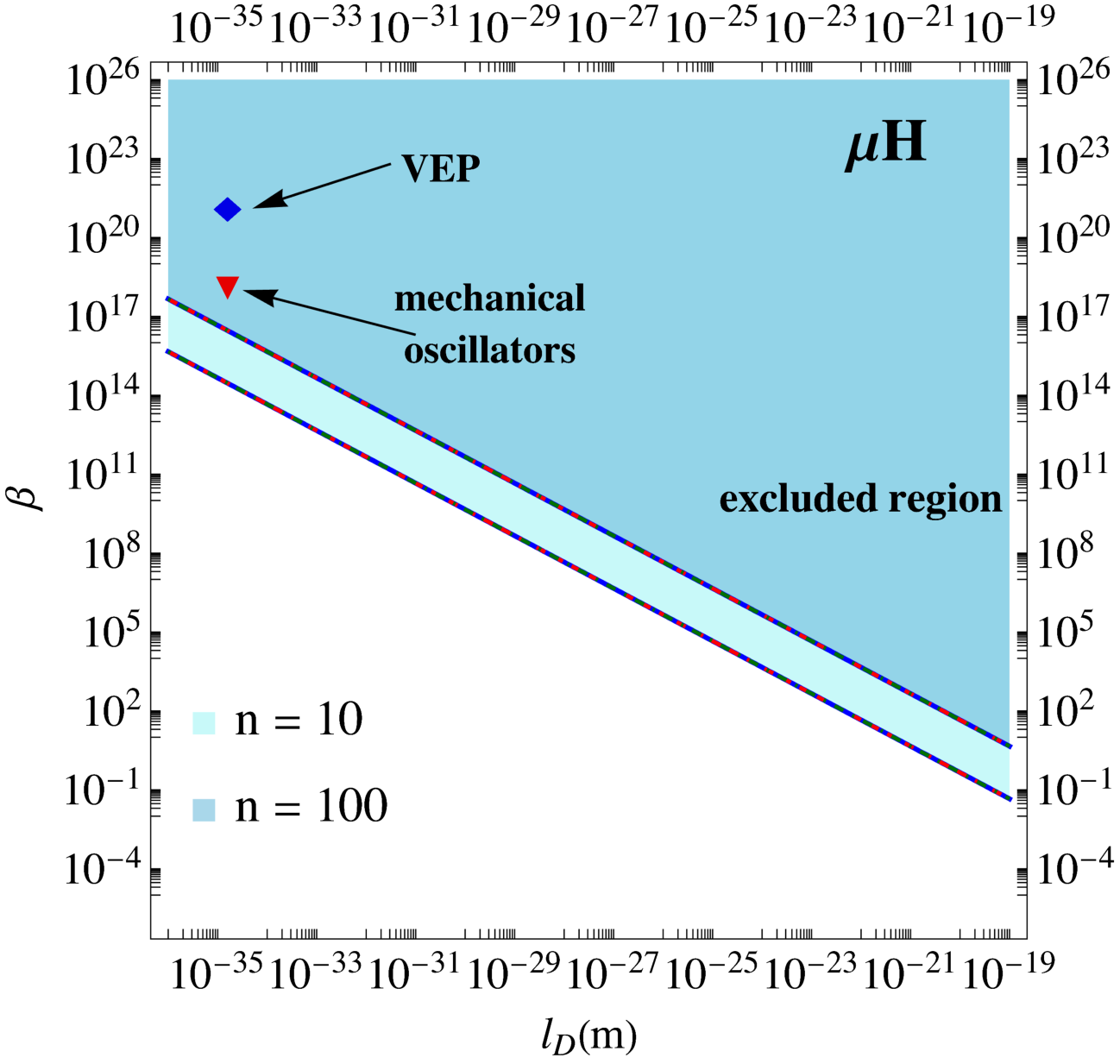}
\par\end{center}
\begin{center}
(b)
\par\end{center}%
\end{minipage}\caption{\label{fig3}Prospective bounds for $\beta$-parameter
and the $D$-dimensional Planck length are obtained from fine structure
constant measurements by considering the muonic hydrogen-like atom
assuming (a) $y=10^{-6}$ and (b) $y=10^{6}$.}
\end{figure}

By specifying the principal quantum number $n$ of a hydrogen-like
atom, we have found upper limits to deformation parameter $\beta$
and Planck length $l_{D}$ in higher-dimensional space. Moreover,
the constraints obtained from experiments of gravitational and non-gravitational
sources also are shown in Figures \ref{fig2} and \ref{fig3}, by
considering electronic and muonic hydrogen-like atoms, respectively.
In these figures, the shaded areas represent excluded regions by our
analysis. The ``VEP" point corresponds to constraint found from
violation of equivalence principle on Earth, while the marker labeled
``mechanical oscillator" describes bound obtained from a non-gravitational
experiment studying opto-mechanical interaction between micro-and
nano-oscillators \citep{KANAZAWA2019}.

As we can see, the constraints found by our analysis are more stringent
than those obtained by other physical systems from gravitational and
non-gravitational sources \citep{KANAZAWA2019}. For instance, for
the gravitational experiment, strong constraints have been extracted
from the analysis of violation of the equivalence principle, where
it was shown that $\beta<10^{21}$ \citep{GHOSH2014}. In another
way, considering constraints derived from micro-and nano-harmonic-oscillators,
the constraint $\beta<10^{18}$ was imposed to the deformation parameter
\citep{BAWAJ2015}. If we assume that the fundamental scale is $l_{D}=1.6\times10^{-35}\mathrm{\mathrm{m}}$
-- for instance, keeping fixed the values $n=10$, $y=10^{2}$, and
$\delta=5$ --, our analysis gives us the restriction to the deformation
parameter $\beta<5.3\times10^{16}$. While for $l_{D}=10^{-20}\mathrm{\mathrm{m}}$
$\left(\sim\mathcal{O}\mathrm{\left(TeV/\hbar c\right)}^{-1}\right)$
we get that $\beta<85$.

As expected, to small values of $y$, the bounds
present strong dependence with $\delta$. Besides, to higher $\delta$-values,
with $l_{D}>10^{-22}\mathrm{m}$, we found the most stringent constraints
{[}see Fig. \ref{fig2}a{]}. On the other hand, higher $y$-values
weaken the correction coming from the brane thickness and therefore
make the bounds independent of spacetime codimensions {[}see Fig.
\ref{fig2}b{]}. In its turn, for the electronic atom the intermediate
length scale is $\ell=4.2\times10^{-19}\mathrm{m}.$ Below this scale,
we expect the effects of Quantum Gravity to become measurable. We
should emphasize that these constraints on $\beta$ ensure that the
effects from GUP have not manifested themselves until currently in
the fine structure constant measure. 

Recent developments in the field of spectroscopy
have provided motivation for exploring the potential of precise measurements
of optical transitions between Rydberg states as an effective means
of testing QED theory and obtaining a more accurate value for the
Rydberg constant \citep{JENTSCHURA2008}. In this context, some works
address the possibility of measuring deviations to the inverse square
law, being able to detect traces of extra dimensions, or even discussing
new Physics using the spectroscopy of Rydberg states \citep{lemos3,LEMOS2021,JONES2020}.
Somewhat, these studies aiming circumvent the proton radius puzzle
\citep{POHL2013}. Here, the high principal quantum number $n$ ensure
that we considered the Rydberg states. Notice that, for high values
of the principal quantum number, one observed the weakening of the
constraints obtained in our analysis. This result may be associated
with the fact that the lepton orbiting a Rydberg atom will be far
away from the atomic nucleus. In this case, the lepton would feel
the field produced by the atomic nucleus more weakly, inducing a smaller
correction on the fine structure constant.

Although we have obtained bounds from electronic hydrogen-like atoms
data, we must emphasize that experiments involving muonic atoms have
received growing attention recently \citep{EIDES2001,POHL2013,KRAUTH2016,ANTOGNINI2021}.
This is partly due to the possibility of looking at extremely small
regions of the order of the proton radius, allowing the study of the
atomic structure \citep{POHL2013}. In addition, such physical systems
would act as tests for Quantum Electrodynamics (QED) since atomic
physics can be applied to measure the fundamental properties of elementary
particles. At last, the muonic systems can still allow us to seek
physics beyond the SM \citep{POHL2013}. Briefly stated, a muonic
atom can be described as an atom whose lepton, usually an electron,
that orbits the atomic nucleus has been replaced by a muon, which
possesses an electric charge identical to the electron \citep{EIDES2001}.

In this case, if we admit that measurements of the fine structure
constant might be extracted from experiments using muonic atoms, we
expect more stringent bounds than early obtained since $m_{\mu}/m_{e}\simeq206$
and $a_{0}/a_{0\mu}\simeq185$, where $a_{0\mu}$ is the Bohr radius
of the muonic hydrogen-like atom. Thus, as previously discussed, we
might find prospective constraints by assuming that the fine structure
constant has been extracted from measurements into a muonic atom with
the same uncertainty that to electronic atom \citep{MOREL2020}. This
analysis yields Figure \ref{fig3}.

As shown in Figure \ref{fig3}, the bounds from the muonic hydrogen
allows us to find stronger constraints than those obtained from the
gravitational and non-gravitational experiments \citep{KANAZAWA2019}. To obtain these results we have considered $y=10^{-6}$
(Fig. \ref{fig3}.a) and $y=10^{6}$ (Fig. \ref{fig3}.b). Although the lepton is away from the atomic nucleus in the Rydberg
atoms, the gravitational effects will be amplified due to the smaller
Bohr radius of the muonic atom. As we can see by tightest the allowed
region for the deformation parameter in Figure \ref{fig3}, the muonic
system provides us with the strongest prospective constraints than
found in the electronic atom.

\section{\label{new}Intermediate length scale as a regulator effect of new
physics}

\begin{figure}[th]
\includegraphics[scale=0.4]{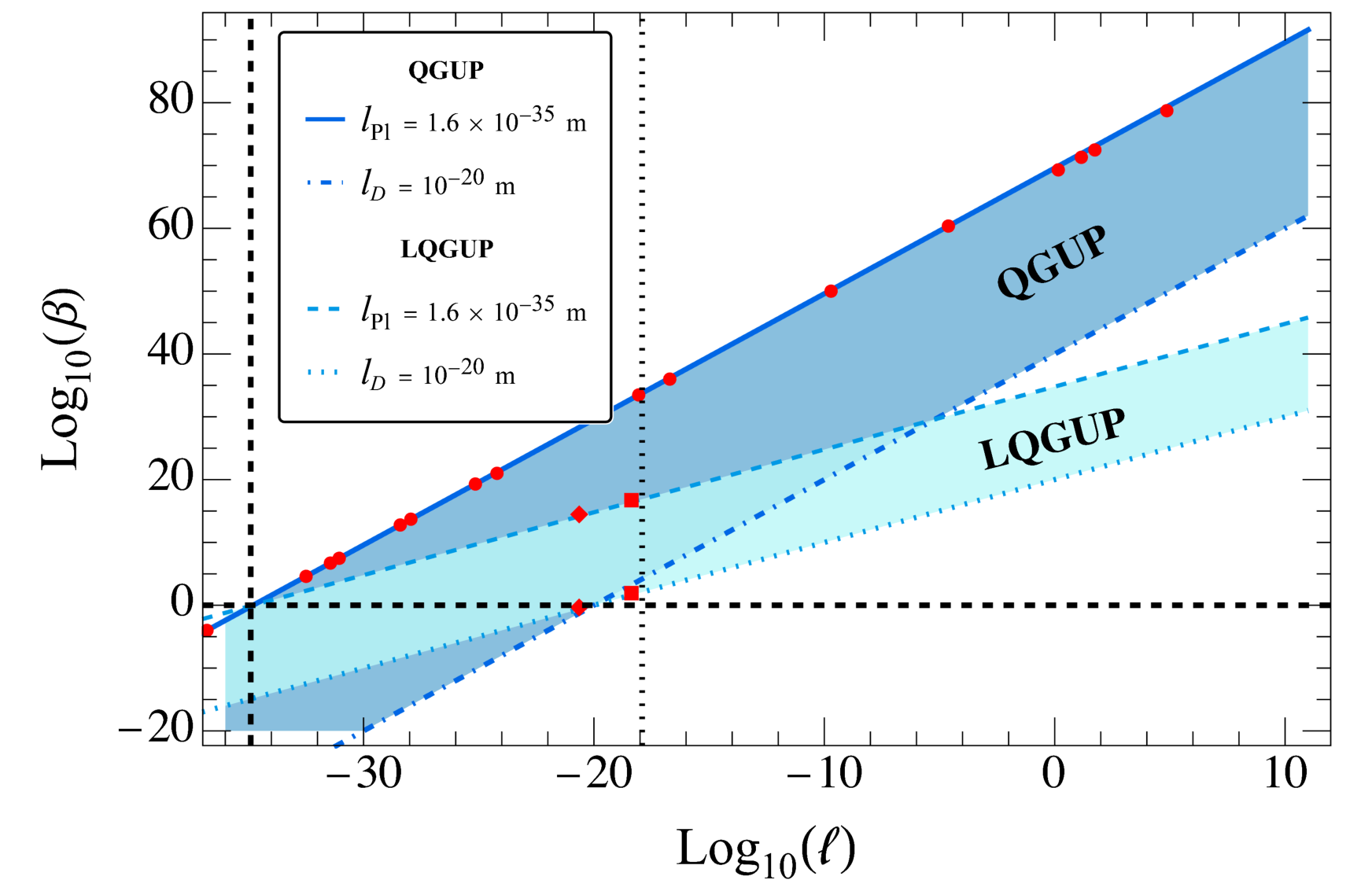}

\caption{\label{fig4}The horizontal black dashed line represents the curve
$\beta=1$, while the vertical black dashed is found for fixed value
$\ell=1.6\times10^{-35}\mathrm{m}$. The vertical dotted line defines
the region below which $\ell\lesssim\mathrm{\mathcal{O}\mathrm{\left(TeV/\hbar c\right)}^{-1}}$.}
\end{figure}

Here we aim to present the constraints on the intermediate length
of large scale $\ell$ and the deformation parameter $\beta$ from
the empirical data. Thus, we consider empirical bounds data on $\beta$
obtained from several physical systems of gravitational and non-gravitational
sources \citep{SCARDIGLI2015,FENG2017,DAS2008,MARIN2013,BAWAJ2015,BUSHEV2019}.
Since the limits on the deformation parameter have been found in the
$3$-dimensional scenario $\left(l_{Pl}=1.6\times10^{-35}\mathrm{m}\right)$,
one can compute the value to intermediate length scale $\ell$ from
these data as shown in Table \ref{tab1}, knowing the predicted deformation
on uncertainty relation. Besides, the constraints have been obtained
by considering quadratic-like GUP corrections (QGUP). In this framework,
one expect that $\beta\sim l_{Pl}^{-2}\ell^{2}$. Indeed, the empirical
data obtained from Table \ref{tab1} are shown in Figure \ref{fig4}
by the red dots and fitted by the solid curve described by the function
$\beta=1.3\times10^{69}\ell^{1.98}$. Thus, for instance, $\beta=1$
leads to $\ell=1.4\times10^{-35}\mathrm{m}$. The dash-dotted curve
is an prediction found by considering $l_{D}=10^{-20}\mathrm{m}$.
We expect that any measurement of Quantum Gravity effects must be
ruled out in the area above the solid curve.

\begin{table}[th]
{\addtocounter{table}{-1}}%
\begin{longtable}[c]{|c|c|c|}
\hline 
$\beta<$ & $\ell\left(\mathrm{m}\right)<$ & Ref.\tabularnewline
\hline 
\hline
\endfirsthead
\hline 
$\beta<$ & $\ell\left(\mathrm{m}\right)<$ & Ref.\tabularnewline
\hline 
\hline
\endhead
\hline 
$5.3\times10^{78}$ & $7.4\times10^{4}$ & \multirow{4}{*}{\citep{SCARDIGLI2015}}\tabularnewline
\cline{1-2} \cline{2-2} 
$3\times10^{72}$ & $5.5\times10^{1}$ & \tabularnewline
\cline{1-2} \cline{2-2} 
$2\times10^{71}$ & $1.4\times10^{1}$ & \tabularnewline
\cline{1-2} \cline{2-2} 
$2\times10^{69}$ & $1.4$ & \tabularnewline
\hline 
$2.3\times10^{60}$ & $2.4\times10^{-5}$ & \citep{FENG2017}\tabularnewline
\hline 
$10^{50}$ & $2.0\times10^{-10}$ & \multirow{3}{*}{\citep{DAS2008}}\tabularnewline
\cline{1-2} \cline{2-2} 
$10^{36}$ & $2.0\times10^{-17}$ & \tabularnewline
\cline{1-2} \cline{2-2} 
$10^{21}$ & $6.2\times10^{-25}$ & \tabularnewline
\hline 
$3\times10^{33}$ & $8.8\times10^{-19}$ & \citep{MARIN2013}\tabularnewline
\hline 
$2\times10^{19}$ & $7.2\times10^{-26}$ & \multirow{4}{*}{\citep{BAWAJ2015}}\tabularnewline
\cline{1-2} \cline{2-2} 
$5\times10^{13}$ & $1.1\times10^{-28}$ & \tabularnewline
\cline{1-2} \cline{2-2} 
$6\times10^{12}$ & $3.9\times10^{-29}$ & \tabularnewline
\cline{1-2} \cline{2-2} 
$3\times10^{7}$ & $8.8\times10^{-32}$ & \tabularnewline
\hline 
$5.2\times10^{6}$ & $3.6\times10^{-32}$ & \multirow{3}{*}{\citep{BUSHEV2019}}\tabularnewline
\cline{1-2} \cline{2-2} 
$4\times10^{4}$ & $3.2\times10^{-33}$ & \tabularnewline
\cline{1-2} \cline{2-2} 
$10^{-4}$ & $1.6\times10^{-37}$ & \tabularnewline
\hline 
\end{longtable}

\caption{\label{tab1}Upper limits to constraints on the deformation parameter
and intermediate length scale from gravitational and non-gravitational
experimental sources.}
\end{table}

On the other hand, the constraints on $\ell$ found by our analysis
(see Eq. (\ref{Intermediate_Length})), which considers empirical
data from the fine structure constant measure, have been labeled by
the marks ``$\blacksquare$" (electronic atom constraints) and ``$\blacklozenge$"
(muonic atom constraints) in the plot. The fitted data by $\beta\left(\ell\right)$
that considers the linear and quadratic GUP (LQGUP) will be represented
in Figure \ref{fig4} by the linear dashed and dotted curves found
keeping the fixed values for $l_{D}=l_{Pl}=1.6\times10^{-35}\mathrm{m}$
and $l_{D}=10^{-20}\mathrm{m}$, respectively. The dashed curve is
plotted from the best fit function $\beta=1.2\times10^{35}\ell$.
In this case, $\beta<1$ implies that $\ell<8.0\times10^{-36}\mathrm{m}$.
In its turn, for predictive dotted curve the function which best fits
the data is $\beta=2\times10^{20}\ell$, so that for $\beta<1$ we
get $\ell<5.0\times10^{-21}\mathrm{m}$.

It is worthwhile to note that the constraints are weakened in $3$-dimensional
scenario $\left(l_{D}=l_{Pl}=1.6\times10^{-35}\mathrm{m}\right)$.
Furthermore, if $l_{D}=l_{Pl}=1.6\times10^{-35}\mathrm{m}$, for any
values of $\ell$ there will not exist $\beta<1$ such that $\ell>l_{Pl}=1.6\times10^{-35}\mathrm{m}$.
Indeed, the experiments with physical pendula have
provided the most stringent limit on the deformation parameter $\beta$
\citep{BUSHEV2019}. However, as one can see, such bounds require
that $\ell<l_{Pl}$. Nevertheless, if the spacetime has hidden dimensions,
there is a wide range to $\ell$-scale so that $\beta<1$ and also
$\ell>l_{Pl}$, although we must have the condition on the intermediate
scale $\ell<l_{D}$. In this context, the existence of a supplementary
space leads to the prospective predictions of bounds $\beta\sim1$.
Thus, by considering the extra-dimensional scenario, the LQGUP strongly
constrains the region such that $\beta\gg1$, while QGUP presents
more stringent constraints for $\beta\ll1$. Traces of new physics
are expected to reveal in the shadowed areas.

\section{Concluding Remarks\label{secIV}}

In this paper, we discuss the generalization of the Heisenberg Uncertainty
Principle (HUP) in thick braneworld background. As
is known, the Generalized Uncertainty Principle (GUP) predicts the
existence of a minimal measurable length. In its turn, in braneworld
scenario, the minimal length is related to the true fundamental scale,
called higher-dimensional Planck length. By considering the hydrogen-like
atoms in Rydberg states within the thick brane framework, we show
that electrostatic potential must be corrected due to gravitational
contribution computable for the electromagnetic interaction. This
correction induces a modification on Coulomb law responsible for describing
the electrostatic interaction between charged particles.

For the sake of simplicity, we address the origin of the correction
for the HUP as being owing to the gravitational interaction effects
at a fundamental length scale. Thus, assuming a combination of linear
and quadratic GUP approaches, we have found the effective fine structure
constant, which depends on the deformation parameter $\beta$ and
$\left(3+\delta\right)$-dimensional Planck length $l_{D}$. By comparing a recent measure of the fine structure constant with the
predicted deviation by our study, we obtain stringent constraints
for the deformation parameter and the higher-dimensional Planck length.
From our analysis -- for instance, by fixing values $n=10$, $y=10^{2}$,
and $\delta=5$ --, if we consider that the fundamental length scale
$l_{D}=l_{Pl}=1.6\times10^{-35}\mathrm{m}$, the upper bound on $\beta<5.3\times10^{16}$,
which avoids any conflicts with the empirical results. On the other
hand, if the true fundamental scale $l_{D}=10^{-20}\mathrm{m}$ $\left(\sim\mathcal{O}\mathrm{\left(TeV/\hbar c\right)}^{-1}\right)$,
this implies in a strengthening of bounds, so that $\beta<85$. From
these results, we compute the intermediate length scale that must
regulate the quantum effects of gravity $\ell=4.2\times10^{-19}\mathrm{m}$
for the electronic atom. We must still emphasize that, in this analysis,
the precision of $\alpha$-measurement is related to the tightening
of bounds.

For atoms that lie in higher states with $n\gg1$, it becomes difficult
to probe regions where the supposed effects of gravitational interaction
are amplified. However, the smallest Bohr radius for the muonic system
leads to the strongest constraints than those found for electronic
hydrogen. From this analysis, we present prospective bounds by considering
Rydberg muonic atoms in the Figure \ref{fig3}, which have approximately
intensified the constraints by a factor of $10^{2}$ times. Therefore,
muonic atoms have shown to be as possible physical systems capable
of testing, simultaneously, GUP and extra-dimensional theories.

Finally, we have discussed the definition of an intermediate length
scale that would act as a regulator of the effects of Quantum Gravity.
We showed that the extra-dimensional scenario leads to constraints
such that $\beta\sim1$ and, notwithstanding the GUP effects may be
measurable. If GUP deviations are detected, in principle, we may infer
the bounds for the higher-dimensional Planck length, which would indicate
the existence of extra dimensions. Although
we have obtained strong constraints on the deformation parameter and
Planck length of the higher-dimensional space, we must highlight that,
in our analysis, such bounds on $\beta$ and $l_{D}$ were obtained
by keeping the other free parameters fixed. In this case, in a sense,
this approach presents prospective constraints get from the investigation
of the space of parameters.
\begin{acknowledgments}
We would like to thank CNPq, CAPES and CNPq/PRONEX/FAPESQ-PB (Grant
no. 165/2018), for partial financial support. ASL and FAB acknowledge
support from CNPq (Grant nos. 150601/2021-2, 312104/2018-9). ASL acknowledge
support from CAPES (Grant no. 88887.800922/2023-00). 
\end{acknowledgments}


\begin{thebibliography}{99}
\bibitem{add1}N. Arkani-Hamed, S. Dimopoulos and G. R. Dvali, Phys.
Lett. B \textbf{429}, 263 (1998).

\bibitem{add2}I. Antoniadis, N. Arkani-Hamed, S. Dimopoulos and G.
R. Dvali, Phys. Lett. B \textbf{436}, 257 (1998).

\bibitem{rs1}L. Randall and R. Sundrum, Phys. Rev. Lett. \textbf{83},
3370 (1999).

\bibitem{rs2}L. Randall and R. Sundrum, Phys. Rev. Lett. \textbf{83},
4690 (1999).

\bibitem{RUBAKOV1983}V. Rubakov and M. Shaposhnikov, Phys. Lett.
B \textbf{125}, 136 (1983).

\bibitem{CSAKI2000}C. Csaki, J. Erlich, T. J. Hollowood and Y. Shirman,
Nucl. Phys. B \textbf{581}, 309-338 (2000).

\bibitem{lemos1}F. Dahia and A. S. Lemos, Phys. Rev. D \textbf{94},
084033 (2016).

\bibitem{lemos2}F. Dahia and A. S. Lemos, Eur. Phys. J. C \textbf{76},
435 (2016).

\bibitem{lemos3}F. Dahia, E. Maciel and A. S. Lemos, Eur. Phys. J.
C \textbf{78}, no.6, 526 (2018).

\bibitem{PALMA2003}G. A. Palma, P. Brax, A. C. Davis and C. van de
Bruck, Phys. Rev. D \textbf{68}, 123519 (2003).

\bibitem{AGUILAR2003}P. Loren-Aguilar, E. Garcia-Berro, J. Isern
and Y. A. Kubyshin, Class. Quant. Grav. \textbf{20}, 3885-3896 (2003).

\bibitem{WILL2014}C. M. Will, Living Rev. Rel. \textbf{17}, 4 (2014).

\bibitem{AMATI1989}D. Amati, M. Ciafaloni and G. Veneziano, Phys.
Lett. B \textbf{216}, 41-47 (1989).

\bibitem{SCARDIGLI1999}F. Scardigli, Phys. Lett. B \textbf{452},
39-44 (1999).

\bibitem{ADLER1999}R. J. Adler and D. I. Santiago, Mod. Phys. Lett.
A \textbf{14}, 1371 (1999).

\bibitem{CAPOZZIELLO2000}S. Capozziello, G. Lambiase and G. Scarpetta,
Int. J. Theor. Phys. \textbf{39}, 15-22 (2000).

\bibitem{CASADIO2023}R. Casadio, W. Feng, I. Kuntz and F. Scardigli,
Phys. Lett. B \textbf{838}, 137722 (2023).

\bibitem{GARRAY1995}L. J. Garay, Int. J. Mod. Phys. A \textbf{10},
145 (1995).

\bibitem{AMELINO2001}G. Amelino-Camelia, Phys. Lett. B \textbf{510},
255 (2001).

\bibitem{ALI2009}A. F. Ali, S. Das and E. C. Vagenas, Phys. Lett.
B \textbf{678}, 497 (2009).

\bibitem{HOSSEN2013}S. Hossenfelder, Living Rev. Rel. \textbf{16},
2 (2013).

\bibitem{TAWFIK2014}A. N. Tawfik and A. M. Diab, Int. J. Mod. Phys.
D \textbf{23}, no.12, 1430025 (2014).

\bibitem{MAGGIORE1993}M. Maggiore, Phys. Lett. B \textbf{304}, 65-69
(1993).

\bibitem{ANACLETO2018}M. A. Anacleto, F. A. Brito, A. G. Cavalcanti,
E. Passos and J. Spinelly, Gen. Rel. Grav. \textbf{50}, no.2, 23 (2018).

\bibitem{ANACLETO2021}M. A. Anacleto, F. A. Brito, B. R. Carvalho
and E. Passos, Adv. High Energy Phys. \textbf{2021}, 6633684 (2021).

\bibitem{IORIO2020}A. Iorio, G. Lambiase, P. Pais and F. Scardigli,
Phys. Rev. D \textbf{101}, no.10, 105002 (2020).

\bibitem{ANACLETO2014}M. A. Anacleto, F. A. Brito, E. Passos and
W. P. Santos, Phys. Lett. B \textbf{737}, 6-11 (2014).

\bibitem{NICOLINI2010}P. Nicolini, Phys. Rev. D \textbf{82}, 044030
(2010).

\bibitem{TAWFIK2013}A. Tawfik, H. Magdy and A.Farag Ali, Gen. Rel.
Grav. \textbf{45}, 1227-1246 (2013).

\bibitem{BARCA2022}G. Barca, E. Giovannetti and G. Montani, Int.
J. Geom. Meth. Mod. Phys. \textbf{19}, no.07, 2250097 (2022).

\bibitem{NASSERI2005}F. Nasseri, Phys. Lett. B \textbf{618}, 229-232
(2005).

\bibitem{NASSERI2006}F. Nasseri, Phys. Lett. B \textbf{632}, 151-154
(2006).

\bibitem{MARQUES2012}G. de A.Marques and V. B. Bezerra, Int. J. Mod.
Phys. Conf. Ser. \textbf{18}, 25-30 (2012).

\bibitem{SCARDIGLI2015}F. Scardigli, R. Casadio, Eur. Phys. J. C
\textbf{75}, 425 (2015).

\bibitem{FENG2017}Z. W. Feng, S. Z. Yang, H. L. Li, X. T. Zu, Phys.
Lett. B \textbf{768}, 81 (2017).

\bibitem{DAS2008}S. Das and E. C. Vagenas, Phys. Rev. Lett. \textbf{101},
221301 (2008).

\bibitem{MARIN2013}F. Marin, F. Marino, M. Bonaldi, M. Cerdonio,
L. Conti, P. Falferi, R. Mezzena, A. Ortolan, G. A. Prodi and L. Taffarello,
\textit{et al.}, Nature Phys. \textbf{9}, 71-73 (2013).

\bibitem{BAWAJ2015}M. Bawaj, C. Biancofiore, F. Marin, Nat. Commun.
\textbf{6}, 7503 (2015).

\bibitem{BUSHEV2019}P. A. Bushev, J. Bourhill, M. Goryachev, N. Kukharchyk,
E. Ivanov, S. Galliou, M. E. Tobar and S. Danilishin, Phys. Rev. D
\textbf{100}, no.6, 066020 (2019).

\bibitem{GHOSH2014}S. Ghosh, Class. Quantum Gravity \textbf{31},
025025 (2014).

\bibitem{DAS2009}S. Das and E. C. Vagenas, Can. J. Phys. \textbf{87},
233 (2009).

\bibitem{ALI2011}A.F. Ali, S. Das, E.C. Vagenas, Phys. Rev. D \textbf{84},
044013 (2011).

\bibitem{BUONINFANTE2019}L. Buoninfante, G. G. Luciano and L. Petruzziello,
Eur. Phys. J. C \textbf{79}, no.8, 663 (2019).

\bibitem{KANAZAWA2019}T. Kanazawa, G. Lambiase, G. Vilasi and A.
Yoshioka, Eur. Phys. J. C \textbf{79}, no.2, 95 (2019).

\bibitem{SCARDIGLI2020}F. Scardigli, Symmetry \textbf{12}, no.9,
1519 (2020).

\bibitem{SCARDIGLI2017}F. Scardigli, G. Lambiase and E.\textasciitilde Vagenas,
Phys. Lett. B \textbf{767}, 242-246 (2017).

\bibitem{SCARDIGLI2019}F. Scardigli, J. Phys. Conf. Ser. \textbf{1275},
no.1, 012004 (2019).

\bibitem{JENTSCHURA2008}U. D. Jentschura, P. J. Mohr, J. N. Tan and
B. J. Wundt, Phys. Rev. Lett. \textbf{100}, 160404 (2008).

\bibitem{LEMOS2021}A. S. Lemos, EPL \textbf{135}, no.1, 11001 (2021).

\bibitem{JONES2020}M. P. A. Jones, R. M. Potvliege and M. Spannowsky,
Phys. Rev. Res. \textbf{2}, no.1, 013244 (2020).

\bibitem{CVETIC1994}M. Cvetic and A. A. Tseytlin, Nucl. Phys. B \textbf{416},
137-172 (1994).

\bibitem{BRITO2014}F. A. Brito, M. L. F. Freire and W. Serafim, Eur.
Phys. J. C \textbf{74}, no.12, 3202 (2014).

\bibitem{WONG2008}S. S. M. Wong. \textit{Introductory Nuclear
Physics}. John Wiley \& Sons, New York (2008).

\bibitem{TAWFIK2015}A. N. Tawfik and E. A. El Dahab, Int. J. Mod.
Phys. A \textbf{30}, no.09, 1550030 (2015).

\bibitem{GANGOPADHYAY2015}S. Gangopadhyay, A. Dutta and M. Faizal,
EPL \textbf{112}, no.2, 20006 (2015).

\bibitem{VAGENAS2019}E. C. Vagenas, A. F. Ali, M. Hemeda and H. Alshal,
Eur. Phys. J. C \textbf{79}, no.5, 398 (2019).

\bibitem{MOREL2020} L. Morel, Z. Yao, P. Clad\'{e} and S. Guellati-Kh\'{e}lifa,
Nature \textbf{588}, no.7836, 61-65 (2020).

\bibitem{POHL2013}R. Pohl, R. Gilman, G. A. Miller and K. Pachucki,
Ann. Rev. Nucl. Part. Sci. \textbf{63}, 175-204 (2013).

\bibitem{EIDES2001}M. I. Eides, H. Grotch and V. A. Shelyuto, Phys.
Rept. \textbf{342}, 63-261 (2001).

\bibitem{KRAUTH2016}J. J. Krauth, M. Diepold, B. Franke, A. Antognini,
F. Kottmann and R. Pohl, Annals Phys. \textbf{366}, 168-196 (2016).

\bibitem{ANTOGNINI2021}A. Antognini, F. Kottmann and R. Pohl, SciPost
Phys. Proc. \textbf{5}, 021 (2021).
\end{thebibliography}
\end{document}